# *The laminar-turbulent transition in a fibre laser*

E.G. Turitsyna[1], S.V. Smirnov[2], S. Sugavanam[1], N. Tarasov[1], X. Shu[1], S.A. Babin[2,3], E.V. Podivilov[2,3], D.V. Churkin[1,3], G. Falkovich[4], and S. K. Turitsyn[1,*]

[1]Aston Institute of Photonic Technologies, Aston University, Birmingham, B4 7ET,UK

[2] Novosibirsk State University, Novosibirsk, 630090 Russia

[3]Institute of Automation and Electrometry, Siberian Branch, Russian Academy of Sciences, Novosibirsk, 630090, Russia

[4]Weizmann Institute of Science, Rehovot 76100 Israel and Institute for Information Transmission Problems, Moscow, 127994 Russia

**Studying transition to a highly disordered state of turbulence from a linearly stable coherent laminar state is conceptually and technically challenging and immensely important, e.g. all pipe and channel flows are of that type[1-2]. In optics, understanding how systems lose coherence with increase of spatial size or excitation level is an open fundamental problem of practical importance[3-5]. Here we identify, arguably, the simplest system where this classical problem can be studied: we learnt to operate a fibre laser in laminar and turbulent regimes. We show that laminar phase is an analogue of a one-dimensional coherent condensate and turbulence onset is through a spatial loss of coherence. We discover a new mechanism of laminar-turbulent transition in laser operation: condensate destruction by the clustering of dark and grey solitons. This is important both for the design of devices exploiting coherent dynamics and for conceptually new technologies based on systems operating far from thermodynamic equilibrium.**

Nature does not allow us to increase the size of a system without eventually losing coherence. For example, in a classical hydrodynamic problem even though a coherent laminar flow through a pipe is always linearly stable, increasing the pipe diameter or the flow rate eventually makes the flow turbulent, vastly increasing drag[1,2,6-9]. To understand and control turbulence, one needs to identify the building blocks of turbulence onset. For a linearly stable system it is a formidable task[2], since we lack the guide of a linear instability analysis, which shows what destroys the laminar state and helps to identify the patterns that appear instead[6-9].

In fibres with normal dispersion, a coherent monochromatic wave or spectrally narrow packets are linearly stable with respect to the modulation instability[10]. Theoretically, in a laser cavity with normal dispersion preventing modulation instability, it could be possible to achieve a classical wave condensation forming a coherent state[11], though a nonlinear four-wave-mixing leading to wave de-phasing is a major challenge. A kinetic condensation of classical waves has been recently observed in two-dimensional optical Hamiltonian systems[12] making an interesting link to the Bose–Einstein condensation[13,14], including condensation of photons[15]. Contrary to the popular perception of laser as a boringly coherent system, operational regimes in many fibre lasers correspond to very irregular light dynamics and low degree of coherence. A quasi-CW fibre laser normally generates so many modes (up to $10^6$), that fluctuations in their amplitudes and phases result in stochastic radiation, which calls for a description in terms of wave turbulence[16-20]. To establish conditions of the existence of a coherent condensate state and to reveal mechanisms of losing coherence in fibre lasers and turbulence onset it is critically important to comprehensively study a laminar-turbulent transition in a fibre laser radiation, as it is done in classical hydrodynamics experiments[2].

In our experiments, increasing the cavity length or the power of a fibre laser, we pass from a coherent laminar state (coherent over the whole time window *t,* having a narrow spectrum and low level of fluctuations) to a turbulent one. Having a laminar-turbulent transition in an optical system allows us to ascend the higher level of generality and to address fundamental questions of non-equilibrium studies in laser context: What are the mechanisms of losing coherence in fibre lasers? Is the transition due to an increase of temporal or spatial complexity?

The experimental set-up is described briefly in Methods. Here we explain what are "space" and "time" in our optical system to allow for meaningful comparison between fibre optics and hydrodynamics on that matter. The radiation is measured in a single point as a function of time. As light makes round trips in the resonator, we measure the radiation within the series of time windows separated by the round trip time $\tau_{rt}$. The result is a function of two variables: a continuous one within a window denoted *t* and a r discrete one, $T = n \times \tau_{rt}$, where *n* is the number of round trips. The fastest process is the linear wave propagation with the speed of light *c*, so that *t*-dependence by the transform *t-x/c* represents the dependence on the spatial coordinate *x* along the resonator. The spectra of the radiation are obtained by Fourier transform over *t*. Energy pumping, dissipation, dispersion and nonlinearity lead to a slow evolution of the spectra over many round trips *T*. In this way, the slow evolution coordinate *T* has the meaning of *time*, while the fast time *t* is equivalent to *the spatial coordinate x*.

To observe the transition from laminar to turbulent regime, we change the laser power. Laminar lasing is realized at a low pump power, turbulent – at a high power. There is a sharp transition in the properties of the laser radiation upon the increase of the power. The optical spectrum width Γ increases almost twice after the power increases by 1% only (Fig. 1a). The total number of generated modes, $N = \Gamma \times (2Ln/c) \sim 10^5$, so the laminar state is fundamentally different from a single-frequency (single longitudinal mode) generation in lasers[21,22]; here *n* is the refractive index. Similar sharp transition at the same power happens with the most probable intensity (Fig. 1b). Below the transition, the generation is quite stable, the intensity fluctuations are small and the intensity probability density function (pdf) has a sharp narrow peak (Fig. 1b, inset) centred at the mean intensity as it should be for a coherent state. Right before the transition, the peak widens a bit but the most probable intensity is still the mean intensity. At the transition, the most probable intensity falls almost twice, while the pdf changes the form and develops the wide approximately exponential tail that manifests a significant probability of high-intensity fluctuations. The transition is also detected as a drop in the background level of the intensity autocorrelation function from a coherent state level to a more stochastic regime (Fig. 1c). Intensity time traces just before and after the laminar-turbulent transition are shown in Fig. S2.

The transition observed corresponds to the loss of coherence in the system. Indeed, spatio-temporal dynamics of radiation is very different in laminar and turbulent regimes. Rather small fluctuations before and recurring spatio-temporal patterns after the transition are seen in Fig.2, a,b. We detect long-living propagating intensity minima both on a stable laminar background (Fig. 2a) and on a strongly fluctuating turbulent background (Fig. 2b). As the typical nonlinear length $L_{NL} = 1/(\gamma I) \sim 1$ km for the transition power, these structures live ~100 nonlinear lengths and, therefore, are coherent. The temporal width of the coherent structures is at the limit of our experimental resolution.

To resolve internal assembly of the observed coherent structures, we use numerical modelling based on generalized scalar Nonlinear Schrödinger Equation (NSE)[10], for more details see Supplementary Information.NSE shares initials with the vector Navier-Stokes Equation describing the fluid flow. While these two models are comparable in their universality, deceptive simplicity and sheer beauty, the former is much more amenable to numerical treatment. NSE was used before to describe coherent structures[23,24], stochastic driven processes in optical fibres[25,26] as well as fibre lasers[11,27,28].

In the specific case of our fibre laser, modelling demonstrates the same laminar-turbulent transition at comparable levels of pump power and remarkably generic behaviour for varying cavity lengths. Moreover, the numerical simulations provide the proof that the laminar state is a coherent condensate and the transition is condensate destruction. Indeed, a coherent condensate must support long acoustic waves with Bogoliubov dispersion relation[29], $\omega \propto k$, in distinction from usual dispersive waves with $\omega \propto k^2$ without the condensate. We calculated the spatio-temporal spectrum $I(k,\omega)$ and found that it has maxima along straight lines in a laminar regime and along parabolic lines after the transition (Fig. S4). The calculated spatio-temporal dynamics reveals coherent structures similar to those observed experimentally (Fig.2, c,d). Remarkably, their internal structure and phase shifts are well-described by the analytical form of a dark and grey solitons (Fig. S5). Note that dark and grey solitons are one of analytical solutions of the one-dimensional NSE[30].

Most important, modelling allows one to reveal the underlying mechanism of laminar-turbulent transition in much detail and show that the solitons are the building blocks of turbulence onset. With an increase of the pump power or the cavity length, more and more solitons are generated, which, at some point, leads to turbulence. Fig. 3 shows how the transition develops over the evolution coordinate *T* at a fixed power. The moment of transition is clearly seen by the spectral widening in Fig. 3a and the breakdown of spatial coherence in Fig. 3b. Solitons proliferate and cluster, creating a deep minimum which breaks the condensate (Fig. 3, c-e). We thus conclude that the laminar-turbulent transition, observed experimentally and modelled numerically, is via the appearance, proliferation and clustering of solitons. In a linearly unstable system solitons may appear as an outcome of instability as, for instance, for capillary wave turbulence[7]. Ours seems to be the first observation of the proliferation of solitons before and during the laminar-turbulent transition in a linearly stable system.

The condensate destruction leads to creation of an intermittent state with still a rather narrow spectrum yet limited spatial coherence (along $t$). In contrast to the traditional (dynamic-system) view that turbulence arises from an increase in the temporal complexity, here the spatial breakdown of coherence is the leading process, similar to a pipe flow[2,9]. Indeed, even when the asymptotic turbulent stage has not yet fully developed, the condensate has already broken into pieces. When the soliton density becomes high and the condensate is filled with deeps and voids, the state is hardly distinguishable from a dense mixture of coherent bright structures (e.g. bright solitons and breathers). Spatio-temporal patterns expand and shrink (having rhombic form in $t$-$T$ plane) with approximately the same velocity as propagation of solitons on them (Fig. 2, b and d). The patterns recur quasi-periodically and move as the whole against the background, which confirms their coherent nature (Fig. S7). The intensity correlation function over the evolution coordinate shows statistical signature of the quasi-periodicity.

Repeating simulations with the only difference in a small initial noise, we find that the laminar-turbulent transition via soliton clustering is stochastic (compare to Ref. 2): the lifetimes of the condensate fluctuate strongly. The probability of surviving a long time decays exponentially as in radioactive decay (Fig. 4), which suggests that after some time theprobability of decay is constant in time and is independent of the time when the laser was excited. To conclude, we observe the laminar-turbulent transition in the fibre laser radiation and reveal a new mechanism of such transition opening new possibilities in studying the classical fundamental problem of turbulence onset in optical devices. We discover the critical role of coherent structures such as dark and grey solitons in destruction of laser coherence making link between soliton theory and turbulence. We anticipate that our results will lead to better understanding of coherence break-up in lasers and development of new optical engineering concepts and novel classes of lasers operating in far-from-equilibrium regimes.

**Methods summary**

The fibre laser used in the experiments has a standard all-fibre design with a cavity made specifically from high normal-dispersion fibre (D=-44ps/nm/km) placed between specially designed all-fibre laser mirrors – fibre Bragg gratings. Mirrors have super-Gaussian spectral profiles of $6^{th}$ order, around 2 nm bandwidth, with dispersion variation less than 10 ps per bandwidth, which is crucial for experimental realisation of the coherent laminar state and the transition to turbulent state (see Supplementary Information for details). Fibre mirrors were written directly in a fibre core using on-site fibre Bragg gratings writing facility following the refractive index longitudinal profile calculated numerically to obtain the desired spectral and dispersion response.

Spectral and temporal properties of the laser radiation were analysed using an oscilloscope of 36 GHz real-time bandwidth being comparable with optical bandwidth of the radiation.

Numerical modelling was based on two complimentary approaches: (a) analysis of longitudinal resonator modes evolution with round trips and (b) computation of field dynamics using generalized NLSEs, see Supplementary Information for details.

**Supplementary Information** is available in the online version of the paper.

**Author Contributions** S.K.T. and G.F. initiated the study, D.V.C. conceived the experiment and carried it out with S.S. and N.T., E.G,T. and S.V.S. designed and conducted the numerical modelling, E.G.T. and X.S. designed special laser mirrors, X.S. fabricated laser mirrors, G.F., S.K.T. and D.V.C. guided the theoretical and experimental investigations, G.F., S.K.T., D.V.C., E.V.P., S.A.B., S.V.S., E.G.T., S.S. and N.T. analysed data, G.F., S.K.T. and D.V.C. wrote the paper.

**Acknowledgements** We thank I. Vatnik for his support at very early stage of experiments. This work was supported by the grants of ISF, BSF and Minerva Foundation in Israel, ERC, Leverhulme Trust, Royal Society in the United Kingdom and the Ministry of Science and Education and Dynasty Foundation in Russia.

**Competing interests statement:** The authors declare no competing financial interests.

Correspondence to: S.K. Turitsyn, e-mail: s.k.turitsyn@aston.ac.uk

**Figures and captions**

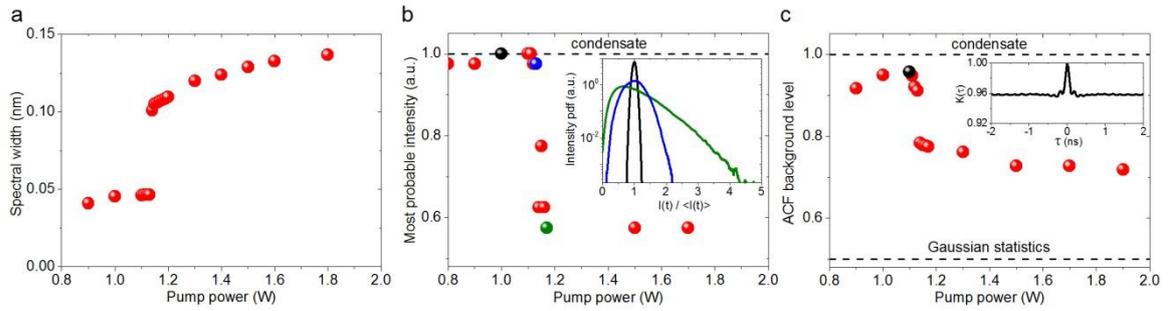

**Figure 1 Laminar-turbulent transition in the fibre-laser experiment**. a) The optical spectrum width (proportional to the number of excited modes) versus power. b) The most probable intensity versus power and the full intensity probability density functions before and after the transition (inset). The colour code attributes curves at the inset to points at the main graph. c) The background level of the intensity autocorrelation function (ACF) $K(\tau) = \langle I(t,T) \times I(t+\tau,T) \rangle$ measured at large $\tau$. Inset shows typical ACF before the transition. For a coherent state, $K(\tau) \to 1$. For a completely stochastic radiation having Gaussian statistics, $K(\tau) \to 0.5$.

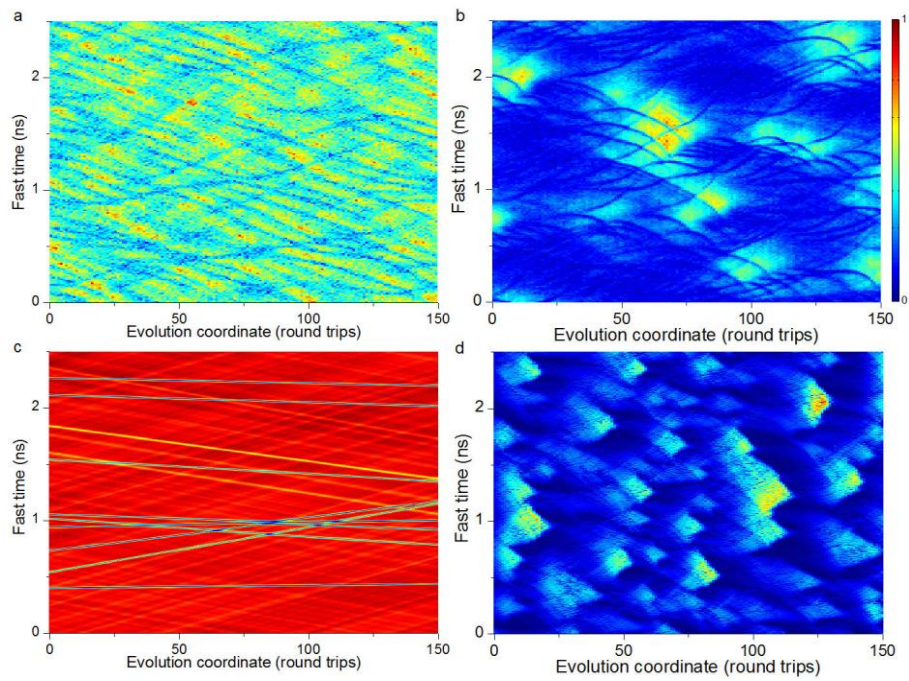

**Fig.2 Coherent structures in spatio-temporal dynamics in experiment and numerical simulation both in laminar and turbulent regimes.** Space-time diagram of the intensity I(t,T) for: a) laminar regime in experiment, b) turbulent regime in experiment, c) laminar regime in modelling, and d) turbulent regime in modelling.

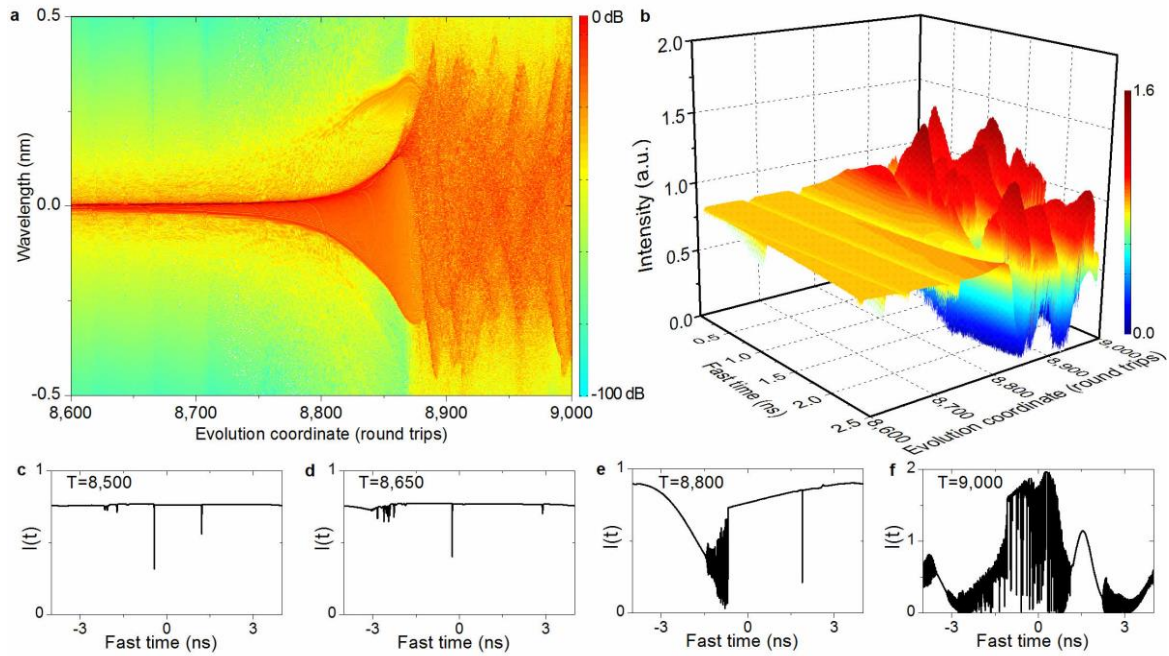

**Figure3 Soliton clustering at laminar-turbulent transition (numeric modelling) at fixed power.** a) Radiation spectrum I(λ) in the logarithmic scale versus T. b) Radiation intensity I(t) versus T: A bunch of solitons creates deepening minimum moving with a negative speed along t (made into a circle i.e. the points t=0 ns and t=2.5 ns are the same). Around T=8,800 round trips, this minimum is deep enough to break the condensate into two pieces, after which the total breakdown into many pieces promptly follows. c-f) Radiation intensity I(t) at four different T: c) condensate with rare isolated solitons, d) beginning of soliton clustering, e) condensate breakdown, f) turbulence. The movie of the whole spectral and spatial evolution is shown in Supplementary Information, Video S1.

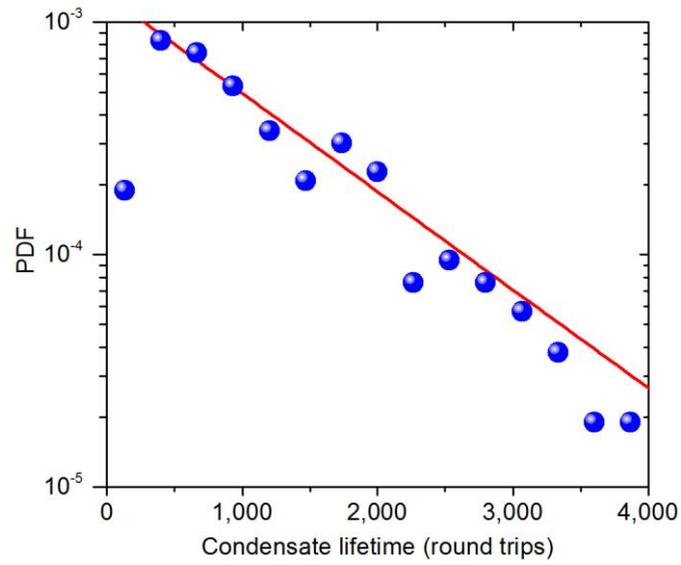

**Figure 4 Probability density function for the condensate lifetime shows probabilistic nature of laminar-turbulent transition via soliton clustering.** Straight line is an exponential approximation at large life times.

**Supplementary Information**

**Experimental details**

Experimentally, we study the transition from laminar to turbulent state in a Raman fibre laser based on high normal dispersion fibre (Fig. S1). We used a standard all-fibre design with a cavity made specifically from high normal-dispersion fibre (D=-44ps/nm/km) placed between specially designed all-fibre laser mirrors – fibre Bragg gratings. The system is pumped at 1455 nm by quasi-CW Watt level pump laser, and lases at Stokes wavelength near 1550 nm. Spectral and temporal properties of laser radiation were analysed using a commercial optical spectrum analyser and an oscilloscope of 36 GHz real-time bandwidth being comparable with optical bandwidth of the radiation.

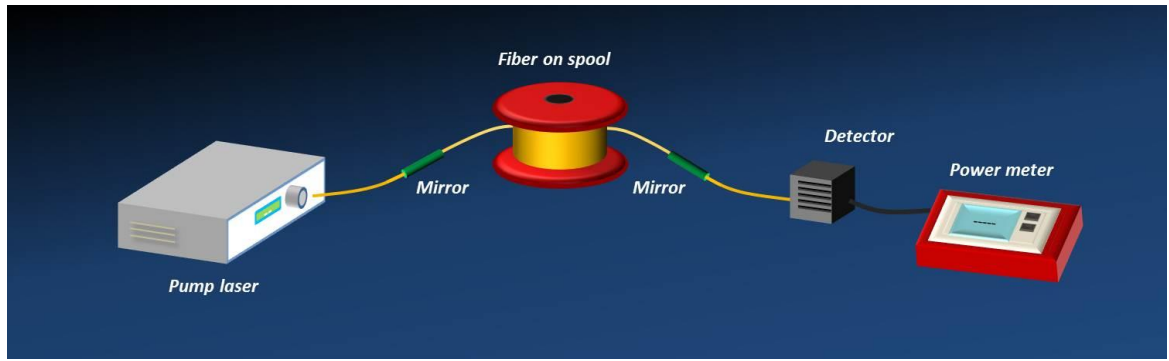

**Figure S1. Experimental setup.**

We measure how the light intensity $I$ depends on the fast time $t$ over a period of time much longer than the laser cavity round-trip time. Each time series could be up to 800 cavity round trips. In the intensity plot $I(t)$, a significant increase of fluctuations is visible at the transition point (Fig. S2). Some fluctuations exist even in the laminar regime because of the residual effects of polarization scrambling, pump wave intensity fluctuations transfer and influence of laser mirrors dispersion (more details below).

Statistical analysis over a number of long time series results in the intensity pdf function (Fig. 1b).

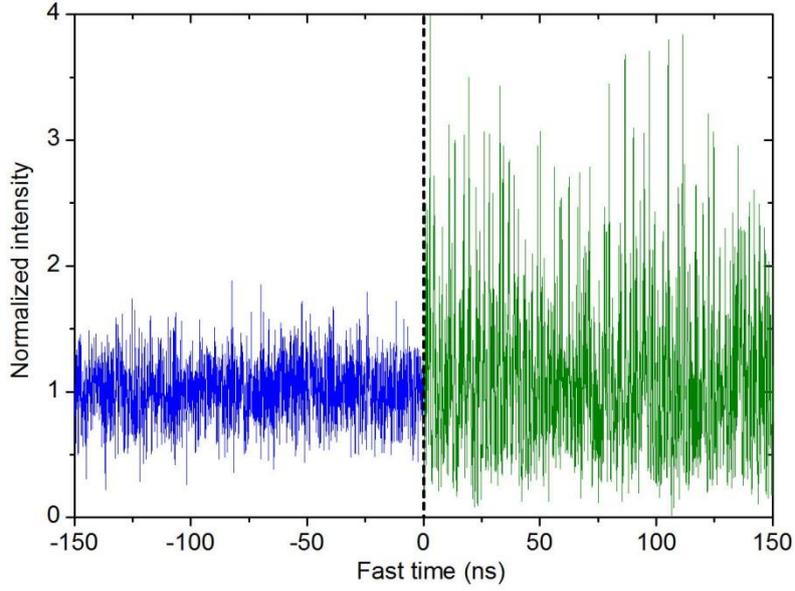

**Figure S2. Experimentally measured laser dynamics before (left, blue) and after (right, green) transition point.** Measurements are done at different moments of time. The colour code and powers are the same as at Fig.1. The fluctuations are suppressed in the laminar regime. The probability of large-intensity fluctuations is higher in the turbulent regime.

To plot the spatio-temporal intensity evolution $I(t,T)$, we first found a cavity round round-trip time $\tau_{rt}$ from the intensity autocorrelation function $K(\tau) = \langle I(t,T) \times I(t+\tau,T) \rangle$ by measuring the time difference between corresponding ACF maxima. Then using $\tau_{rt}$, we slice long time traces $I(t)$ into segments. Each segment has a length equal to round trip time $\tau_{rt}$. Finally, we plot segments one after another using colour code for intensity values. The resulted two dimensional intensity evolution $I(t,T)$ over fast time $t$ and evolution coordinate $T$ (Fig. 2) allows us to analyse spatio-temporal properties of the radiation, including the presence of a long-lived structures.

To establish laminar regime experimentally, the main factors were low level of noise in the system and special laser mirrors. We have observed that an additional noise in the pump wave makes generation turbulent at all powers. Apparently, intensity fluctuations in the pump wave are transferred through cross-phase modulation (XPM) to the generation Stokes wave that leads to stochastization and spectral broadening of the generation wave[1]. Such transfer was observed experimentally in radio-frequency spectrum[2]. In our experiment, we minimized the influence of pump-to-Stokes noise transfer by using the special pump laser with suppressed relative intensity noise down to the value of -120 dB/Hz.

Polarization scrambling also could destroy the laminar regime. In the experiment, we can achieve laminar generation in one polarization by using 2 polarization controllers placed inside the cavity. We use independent polarization measurements to control the state of polarization during measurements. Note that the residual polarization scrambling still exists as we have degree of polarization value up to 0.9 only and DOP could be lower at frequencies above our polarization measurement bandwidth (250 MHz). This could be a source of noise measured in the experiment.

In the experiment realization of laminar-turbulent transition we used specially designed dispersion-free ultra-wideband super-Gaussian shaped laser mirrors. The laser mirror dispersion was found to be crucial for observing the transition. Using mirrors with standard (non-optimized non-zero) dispersion profiles, we always observed only turbulent stochastic generation both in the numerical modelling and in the experiment. In this case, the condensate is destroyed by the phase randomization of the modes after each round trip reflection from the mirrors. So we have used mirrors have super-Gaussian spectral profiles of $6^{th}$ order, with dispersion variation less than 10 ps per 2 nm bandwidth, which is crucial for experimental realisation of transition. Mirrors were written by an on-site Bragg gratings writing facility directly in a fibre core, following the longitudinal profile of the refractive index, calculated numerically to obtain desired spectral and dispersion response.

Finally, the spectral shape of the laser mirrors is also of critical importance for the realization of the laminar generation. There is an energy flux from the condensate to the spectral wings. The wings are filtered out from the cavity at each reflection from the mirror. When mirrors were spectrally narrow or had smooth bell-shaped spectral profiles, this resulted in more energy losses and, therefore, more energy was needed from the condensate to recover the spectral wings at next round-trip. To achieve the generation of a condensate over many round trips, we used the specially designed and fabricated laser mirrors of rectangular wide-band spectral profiles. Fig. S3 shows remarkable dependence of the radiation linewidth on the mirror spectral width in two lasers having laser mirrors of conventional dispersion of ~1000 ps/nm (Fig. S3, blue circles) and dispersion optimized laser mirrors (Fig. S3, red circles). For the laser system with conventional laser mirrors, the generation is always turbulent, and the generation spectral width and, consequently, the number of generated modes increases monotonically with the mirror spectral width (Fig. S3, blue circles), as expected for a usual laser[3]. For the laser system with the dispersion-optimized mirrors, the radiation spectral width is comparable to the mirror spectral width and grows only up to 1.5 nm of mirror spectral width. Here the generation is turbulent. Then, contrary to naïve expectations, the radiation spectral width *decreases* with increasing the mirror spectral width and eventually saturates at low value (Fig. S3, red circles), when the laminar generation is established. Apparently, for a very wide spectral profile of the mirror, there are no losses and the radiation does not change its spectral shape during a roundtrip. For a narrower spectral profiles of the mirror, the losses on spectral wings upon reflection must be compensated by pumping during the roundtrip; the resulting change with *t* leads to spectral broadening.

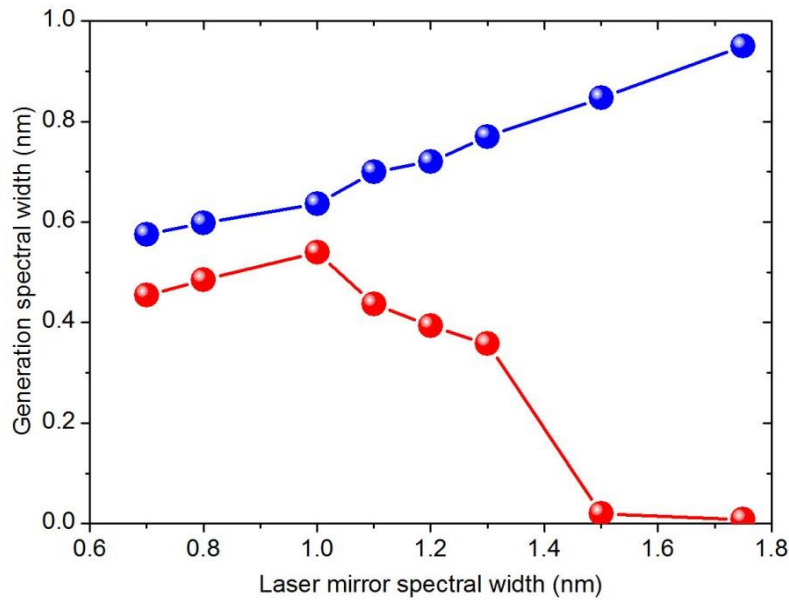

**Figure S3. Numerically calculated laser generation spectral width as a function of the laser mirror spectral width for identical fibre laser systems different only by the laser mirror dispersion.**

Blue circles correspond to the case when the laser mirrors have sufficient dispersion, and the laser generates in turbulent regime only. Red circles correspond to the case when the laser mirrors are dispersion optimized, so the laminar generation is feasible at large spectral width of mirrors.

**Numerical modelling.**

Numerical modelling was based on two complimentary approaches: (a) analysis of longitudinal resonator modes evolution with round trips and (b) computation of field dynamics using generalized NSEs.

Analysis of longitudinal resonator modes is based on the well-established mathematical model that presents the standard slow round trip evolution equation for the longitudinal modes $E_m = \int E(t)e^{-i\Omega_m t}dt$ of the envelope, which can be derived from the generalized Schrödinger equations for backward and forward Stokes waves:

$$\frac{\tau_{RT}}{L}\frac{dE_m}{dT} = \left(G_m - i\beta_2\Omega_m^2\right)E_m - i\gamma\sum_{i,k}E_i E_k E_{i+k-m}^*.$$

Here, $\tau_{rt} = 2Ln/c$ is the round trip time ($n$ is the refractive group index), L is the resonator length and c is the speed of light; the terms on the right-hand side describe, respectively, gain/loss ($G_m$), group-velocity dispersion ($\beta_2$), and the four-wave nonlinear interaction (including cross-phase and self-phase modulation) induced by the Kerr nonlinearity. This equation has the simplest stationary solution in the form of an ideal one-mode (or a pair of modes) condensate that corresponds to the maximum of ($G_m$). Recall that such a monochromatic wave is linearly unstable in the case of ($\beta_2<0$) (the so-called modulation instability) and is linearly stable in the case of ($\beta_2>0$). We consider here the case of normal ($\beta_2>0$) dispersion.

In the simplest scheme, we treat only the integral pumping intensity (the part of the gain/loss $G_m$) as a function of $T$. For more detailed comparison with the experiment and better resolution of the details of t-dependence we used also a more sophisticated model which is the coupled set of NSEs for the radiation and the pumping wave amplitudes as functions of both $T$ and $t$.

$$\frac{\partial E_p^\pm}{\partial T} + \left(\frac{1}{v_s} - \frac{1}{v_p}\right)\frac{\partial E_p^\pm}{\partial t} + \frac{i}{2}\beta_{2p}\frac{\partial^2 E_p^\pm}{\partial t^2} + \frac{\alpha_p}{2}E_p^\pm = i\gamma_p\left|E_p^\pm\right|^2 E_p^\pm - \frac{g_p}{2}\left(\left\langle\left|E_s^\pm\right|^2\right\rangle + \left\langle\left|E_s^\mp\right|^2\right\rangle\right)E_p^\pm$$

$$\frac{\partial E_s^\pm}{\partial T} + \frac{i}{2}\beta_{2s}\frac{\partial^2 E_s^\pm}{\partial t^2} + \frac{\alpha_s}{2}E_s^\pm = i\gamma_s\left|E_s^\pm\right|^2 E_s^\pm + \frac{g_s}{2}\left(\left\langle\left|E_p^\pm\right|^2\right\rangle + \left\langle\left|E_p^\mp\right|^2\right\rangle\right)E_s^\pm$$

Here $t$ stands for time in the frame of references moving with the pump wave, $v_s v_p$ are pump and Stokes waves group-velocities, ± denotes counter-propagating waves, "s" and "p" are used for Stokes and pump waves. At generation power levels less than 1 W, the typical nonlinear length $L_{NL}$ is around 1 km that is much longer that walk-off length $L_w$ ~1 m related with difference of pump and Stokes waves group velocities $v_p$, $v_s$. So equations are averaged over length $L_a$ being $L_w \ll L_a \ll L_{NL}$. Thus we omit XPM term between pump and Stokes waves. The latter model takes into account the fluctuation of the gain because of the fluctuations in highly multi-mode and partially coherent turbulent pump wave. So the resulted condensate in the latter model is noisier than in the model with integral pumping intensity. Except this different of amount of on the condensate, both models provide identical results.

The values of all coefficients used in numerical modelling are: $\alpha_p = 0.07$ dB/km, $\alpha_s = 0.052$ dB/km, $\beta_{2p} = 34$ ps$^2$/km (normal), $\beta_{2s} = 56$ ps$^2$/km (normal), $\gamma_s = 3.5$ (km*W)$^{-1}$ (at pump wavelength) $g_s = 0.6$ (km*W)$^{-1}$ (at pump wavelength), the cavity length was normally a specific L=770 m, but we have also checked the generality of the results by varying the cavity length (e.g. 370 m in Fig. 2 up to 13 km in Fig. 3). We have observed very similar results on transition to turbulent regimes for all systems we considered. Moreover, the results of the two different numerical models agree, the only difference is that the second model demonstrates slightly higher level of fluctuations in the laminar phase.

In addition to the power, spectral properties and intensity evolution of the radiation, in numerical modelling we can calculate properties impossible to measure up to date. Indeed, the main limitation of the experiment is that one measures only intensity values, and all phase information is lost. In numerical modelling, one has both intensity and phase information, which allows, in particular, to calculate the dispersion relation. Once lasing regime is established, temporal distribution of complex field *A(t)* is stored at the beginning of each round-trip large number of successive round-trips. Then two-dimensional fast Fourier transform gives the spatio-temporal spectrum $I(\omega, k)$ plotted as two-dimensional color graph in Fig. S4.

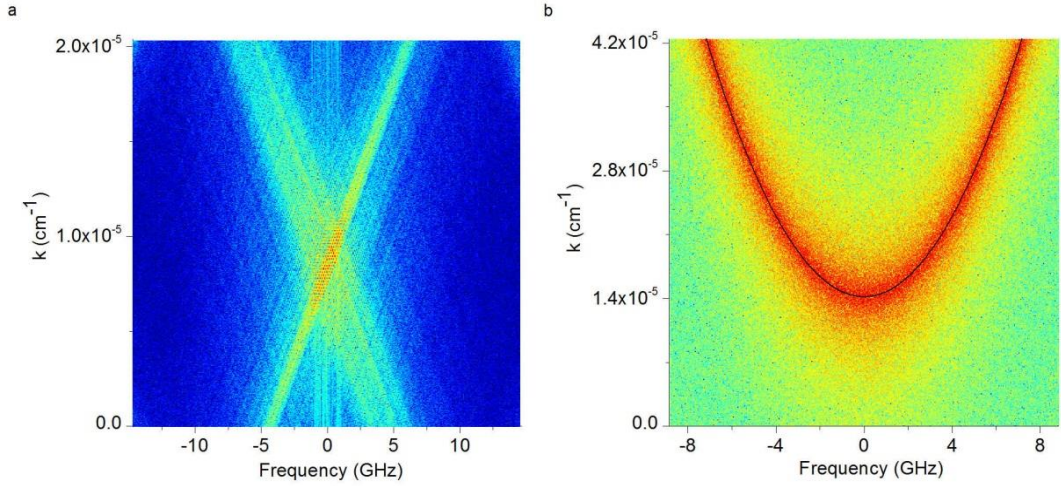

**Figure S4. Numerically calculated dispersion relations in a) laminar and b) turbulent regimes.** Color code is used to display intensity in logarithmic scale. Black solid curve is a parabolic eye guide.

In addition, numerical modelling allows us to resolve the internal temporal shape of the coherent structures, which could not be experimentally measured because of limited real-time resolution. Fig. S5 is devoted to the identification of the coherent structures that proliferate before the transition. Typical coherent structure is shown at Fig. S5a to have a specific dark-soliton $tanh(t)$ shape[4]. The phase shift on a soliton is directly related to the soliton amplitude and is exactly π for a dark soliton. Grey solitons have continuous phase across the structure which is clearly visualized on the phase portrait (fig. S5b). Non-interacting solitons should looks like a chord on the phase portrait. The deeper the soliton, the closer the chord to the ring center should be, so the lines inside the ring demonstrate the phase change i.e. the topological nature of the objects. Bended lines are attributed to interacting solitons .The width of the ring is an amount of the condensate amplitude fluctuations. Fig. S5 proves that, indeed, coherent structures appeared on the condensate in a fibre laser are dark and grey solitons. Numerous such solitons are seen running in Fig. S6. (a view from the bottom – zero intensity level). In particular, the condensate and the initial stage of occurrence of dark and grey solitons that propagate stably and interact with (cross) each other without visible destruction are shown very clearly.

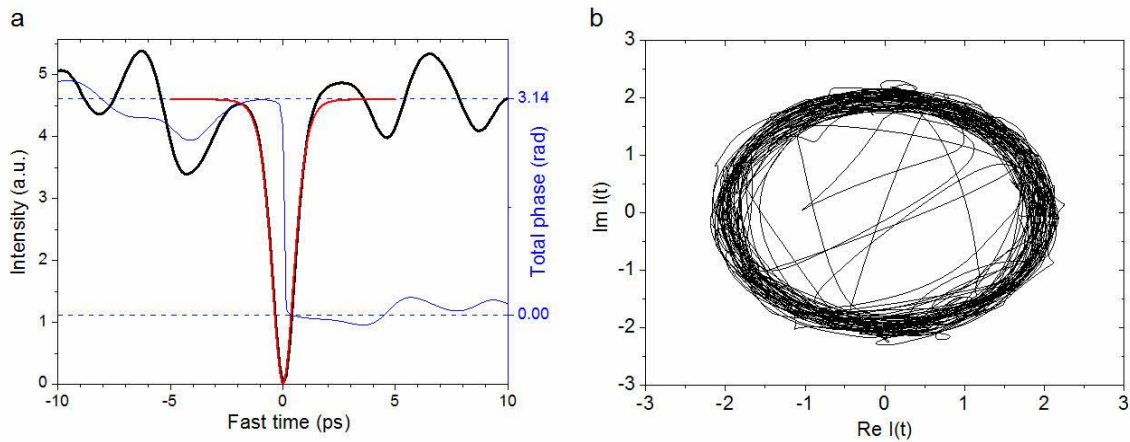

**Figure S5. Dark and grey solitons in the laminar regime.** a) The intensity dynamics *I(t)* taken from numerical simulations (black) in the laminar regime shown with large zoom over fast time *t*. Analytical shape of a dark soliton (red) fits well the coherent structure temporal shape. The numerically calculated total phase (blue) exhibits π-shift over the dark soliton in accordance with dark soliton analytical properties. b) Phase diagram illustrating dark and grey solitons phase shifts.

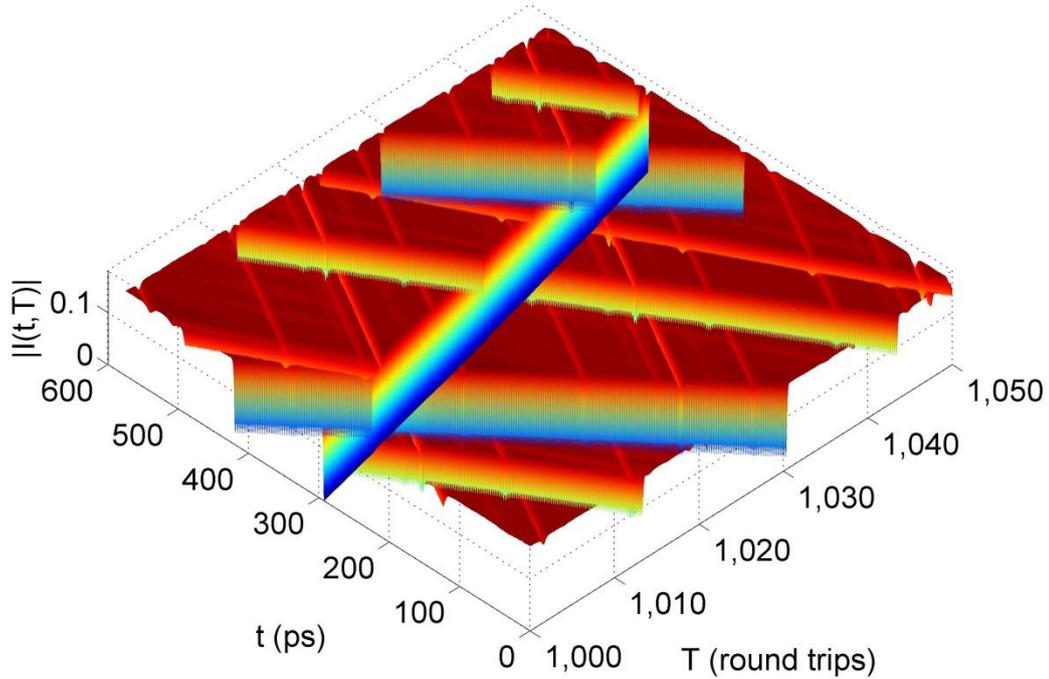

**Figure S6. Solitons running over the condensate.** One can see one dark standing soliton and several grey solitons running in both directions.

To visualize the emergence of dark and grey soliton clusters in all its complexity and highlight their key role in laminar regime destruction, we present a movie, which shows how the intensity evolves over frequency, time and coordinate (Video S1). The upper panel shows the evolution of the spectral intensity $I(\omega,t)$ with the Fourier transform calculated by a short-time Fourier transform with $t$-interval divided over short 256 windows. The lower panel shows the evolution of $I(t)$. One can see solitons appear, proliferate and then cluster creating a minimum (running to the right in this case) which eventually splits the condensate and leads to turbulence. The detailed analysis of the evolution of spectral patterns accompanying soliton clustering will be published elsewhere.

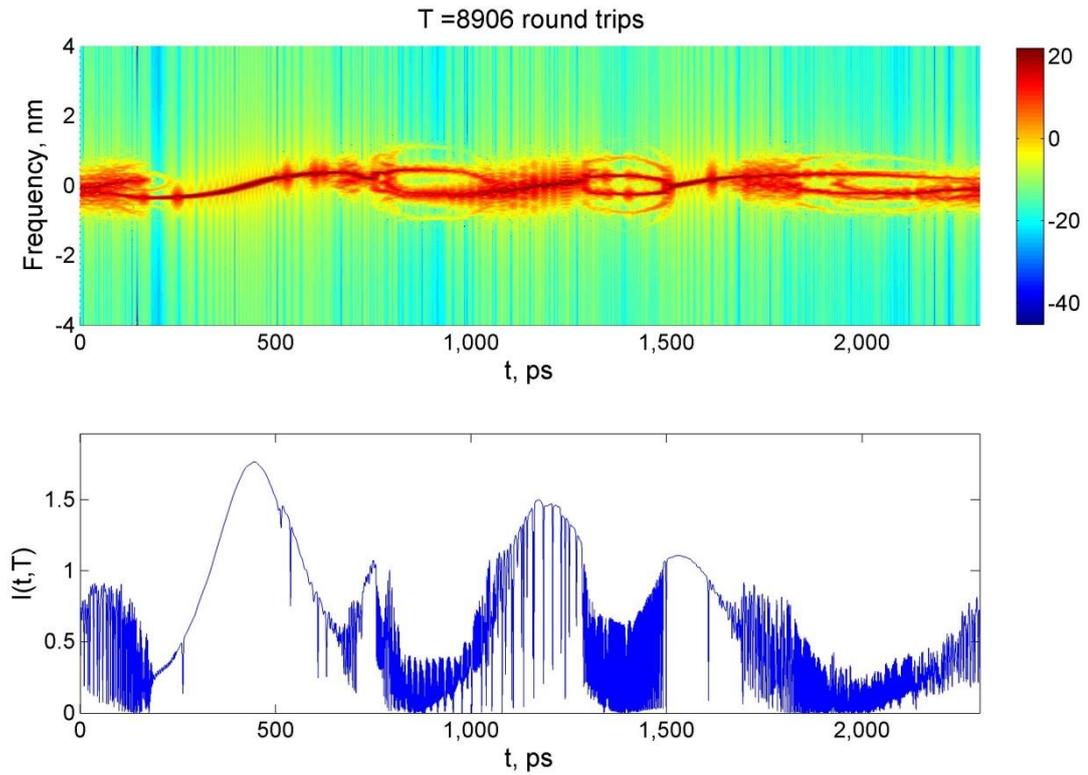

**Video S1. Condensate destruction through clustering of dark and grey solitons.** Upper panel – the radiation spectrogram showing the local frequency spectrum of the part of radiation within a moving 2300 ps temporal window. Lower panel – intensity dynamics $I(t)$. The video is provided in separate file.

After condensate destruction, some order survives in the system in the form of quasi-periodicity, as can be seen from Fig. S7 presenting the dependence of the intensity correlation function on the shift in $T$.

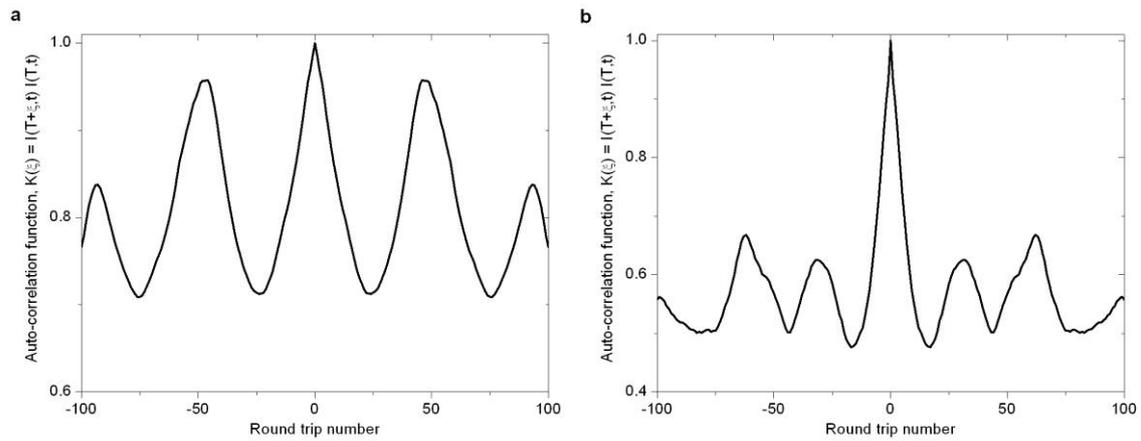

**Figure S7. Intensity autocorrelation function over evolution coordinate**. a) Experiment. b) Numerical simulations.